\documentclass[conference,single]{IEEEtran}
\IEEEoverridecommandlockouts
\usepackage{cite}
\usepackage{amsmath,amssymb,amsfonts}
\usepackage{bm}
\usepackage{algorithmic}
\usepackage{graphicx}
\usepackage{textcomp}
\usepackage{xcolor}
\usepackage{capt-of}  
\usepackage{cuted}    
\usepackage{multirow}
\usepackage{lipsum}
\usepackage{hyperref}
\def\BibTeX{{\rm B\kern-.05em{\sc i\kern-.025em b}\kern-.08em
    T\kern-.1667em\lower.7ex\hbox{E}\kern-.125emX}}
\begin{document}

\title{Multi-Task Learning for Multi-User CSI Feedback\\}

\author{\IEEEauthorblockN{Sharan Mourya}
\and
\IEEEauthorblockN{SaiDhiraj Amuru}
\and
\IEEEauthorblockN{Kiran Kumar Kuchi}
}


\maketitle

\begin{abstract}
Deep learning-based massive MIMO CSI feedback has received a lot of attention in recent years. Now, there exists a plethora of CSI feedback models mostly based on auto-encoders (AE) architecture with an encoder network at the user equipment (UE) and a decoder network at the gNB (base station). However, these models are trained for a single user in a single-channel scenario, making them ineffective in multi-user scenarios with varying channels and varying encoder models across the users. In this work, we address this problem by exploiting the techniques of multi-task learning (MTL) in the context of massive MIMO CSI feedback. In particular, we propose methods to jointly train the existing models in a multi-user setting while increasing the performance of some of the constituent models. For example, through our proposed methods, CSINet when trained along with STNet has seen a $39\%$ increase in performance while increasing the sum rate of the system by $0.07bps/Hz$.
\end{abstract}


\section{Introduction}
With the introduction of Deep Learning techniques \cite{csinet}, channel state information (CSI) feedback for frequency division duplex (FDD) systems has seen an incredible increase in performance and a sizeable reduction in complexity and power consumption. CSINet \cite{csinet} was the first to introduce an  auto-encoder (AE) \cite{vae} based on Convolutional Neural Networks (CNNs) to the compression problem. CSINet employed a fixed receptor size for the CNN to capture the spatial correlation in the angular-delay domain. In \cite{crnet}, a similar approach with multiple receptor sizes was proposed. CLNet\cite{clnet} introduced a simple technique to combine the real and imaginary values of a channel matrix, unlike other models that treated them separately. STNet \cite{stnet} proposed a transformer \cite{transformer} network that has state-of-the-art performance with $1/10^{th}$ resource utilization of other transformer models.\par
All these neural networks are auto-encoder models with an encoder network at the user equipment (UE) and a decoder network at the gNB (base station). The encoder compresses the channel matrix by a factor, which is called the\textit{ compression ratio} and transmits it to the decoder to reconstruct the channel matrix. This encoder-decoder model is then trained in a single-user scenario with either indoor or outdoor channel conditions. This trained model only works in that specific scenario, which is often not useful in reality as the gNB addresses many users with different channel conditions at once. Thus, a model trained for only a specific scenario cannot be employed as a general solution. In addition, there exist numerous CSI feedback models with varying complexities and processing capabilities. Thus, in a multi-user scenario, every user could use a different CSI feedback model whichever meets their requirements. Consequently, the decoder network at gNB has to support a variety of encoder networks to perform well. This is where multi-task learning (MTL) is essential wherein a single model can be trained with multiple datasets i.e, multiple channel scenarios or multiple auto-encoder models.
\par
In \cite{past}, Li et al. proposed multi-task learning techniques for a multi-scenario use case where users are experiencing different channel conditions. However, they restricted the users to having the same encoder architecture to transmit CSI feedback. This limitation is not practical as different users employ different architectures depending on their power and memory requirements. In this letter, we address this issue by allowing users to have different encoder architectures as well as different channel conditions. They also designed a new neural network to address the multi-scenario case whereas we propose to use MTL techniques with existing high-performance models, making our approach generalizable to any model. Zhang et al. \cite{diff} worked on reducing the memory and training time requirements of the existing CSI feedback models by using multi-task learning. Nevertheless, their approach was applicable only to a single-user and single-channel scenario. In contrast, we propose a methodology to use the existing CSI feedback models for multi-user and multi-scenario conditions while reducing resource utilization.  

\section{Motivation}
In the context of CSI feedback, user distributions in a cell can be broadly classified into
\begin{enumerate}
    \item \textit{Single-Scenario Single-Model (SSSM)}: Each user is experiencing the same channel conditions and all are using the same encoder model.
    \item \textit{Single-Scenario Multi-Model (SSMM)}: All users are experiencing the same channel conditions but each user is using a different encoder model.
    \item \textit{Multi-Scenario Single-Model (MSSM)}: Each user is experiencing different channel conditions but all are using the same encoder model.
    \item \textit{Multi-Scenario Multi-Model (MSMM)}: Each user is experiencing different channel conditions and is using a different encoder model.
\end{enumerate}
Users in a cell can be grouped according to these distributions and in reality, gNB might be dealing with more than one user distribution at once. In all these cases, gNB has to decode the CSI feedback from varied user distributions but the decoder is only trained for the SSSM case. The usage of this decoder in the other three cases results in a significant drop in performance. For example, consider three CSI feedback models CSINet\cite{csinet}, CLNet\cite{clnet}, and STNet\cite{stnet} trained for indoor channel scenarios at a $1/4$ compression ratio. We then paired CSINet's encoder with the decoders of CLNet and STNet and vice versa. Table I shows the normalized mean-squared error (NMSE) \cite{csinet} performance of each encoder model paired with every other decoder model on a COST 2100 dataset \cite{cost}.

\begin{figure}[h]
\centering
\includegraphics[width=3in]{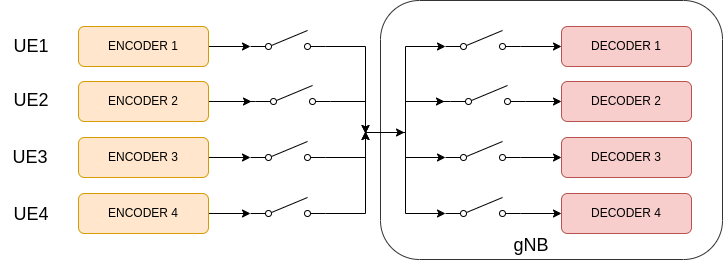}
\captionof{figure}{gNB with multiple decoders. Here, four users have four different encoders whose corresponding decoders are available at the gNB. So, whenever the gNB serves a user, it automatically switches to the corresponding decoder.}
\label{switch}
\end{figure}

\begin{table}[h!]
\caption{NMSE at 1/4 compression ratio and indoor channel conditions}
\centering
\begin{tabular}{|c|c|c|c|} 
\hline
 Model & CSINet Decoder& CLNet Decoder & STNet Decoder\\
 \hline
 CSINet Encoder & -17.36 & 14.53 & 15.56 \\ 
 \hline
 CLNet Encoder& 13.52 & -29.16 & 16.31 \\
 \hline
 STNet Encoder & 10.46 & 15.13 & -31.81 \\
 \hline
\end{tabular}\\
\end{table}
From Table I, we see that the performance degrades when not using the right encoder-decoder pair. Although all three models are trained for exactly the same scenario, they seem to have no correspondence between them. This is more profound if we use the model trained for one channel condition to evaluate a different channel condition. Table II lists the NMSE values of CSINet trained in one scenario and evaluated in another scenario. Performance degradation is evident where ever the right scenario is not evaluated.
\begin{table}[h!]
\centering
\caption{NMSE at 1/4 compression ratio for CSINet}
\begin{tabular}{|c|c|c|} 
\hline
 Model & CSINet (Indoor) & CSINet (Outdoor)\\
 \hline
 CSINet (Indoor) & -17.36 & 13.29\\ 
 \hline
 CSINet (Outdoor)& 11.74 & -8.75\\
 \hline
\end{tabular}\\
\end{table}
Tables I and II indicate that a single model cannot be employed to decode CSI feedback in multi-user and multi-scenario conditions. One way to mitigate this is shown in Fig.{~\ref{switch}} where we use multiple decoders at gNB each corresponding to a different auto-encoder model and switch between these depending on the channel condition and the encoder that the user has employed, which can be identified by a specifier sent along with CSI feedback. This not only increases the feedback overhead but also makes the gNB's design more complicated. This approach is also not scalable as the number of decoders at gNB grows proportionally with the number of different encoders used by users in the cell. 
\par
\begin{figure}[h]
\centering
\includegraphics[width=3in]{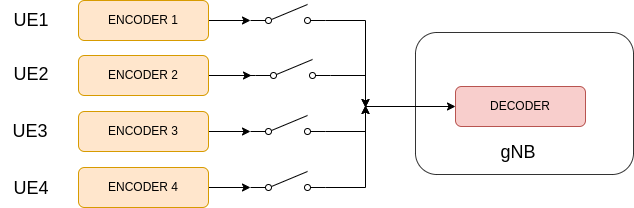}
\captionof{figure}{gNB with a single decoder. Here, four users have four different encoders and a single decoder has to be trained to decode the CSI feedback from all the encoders.}
\label{one}
\end{figure}

 So the need for generalizing the CSI feedback model arises where ideally the gNB uses a single decoder model to decode CSI feedback from varied user distributions as shown in Fig.{~\ref{one}}. One way to solve this problem is by using multi-task learning (MTL)\cite{mtl} where we define each scenario and encoder model as a task and train the gNB's decoder with all the tasks together to decrease the complexity of the gNB.
For the purpose of this study, we consider two channel scenarios: indoor and outdoor, and two different auto-encoder models: CSINet, and STNet. However, the concepts in this paper can be generalized to any number of models and scenarios. Now, each user has two attributes: \{\textit{Scenario, Encoder Model}\}, and those users with the same attributes fall under the same task.

\section{System Model}

Let's consider an orthogonal frequency division multiplexing (OFDM) system with $\Tilde{N}_{c}$ sub-carriers, and $N_{t}\times1$ antenna configuration i.e., the base station (gNB) has $N_{t}$ antennas and the user equipment (UE) has 1 antenna. In this system, the signal on the $n^{th}$ sub-carrier is denoted by
\begin{equation}
    y_{n} = \textbf{$\bm{\Tilde{h}_{n}}^{H}$}\bm{v_{n}}x_{n} + \textbf{$w_{n}$},
\end{equation}
where, $\bm{\Tilde{h}_{n}} \in \mathbb{C}^{N_{t}\times 1}$ is the channel, $\bm{\Tilde{v}_{n}} \in \mathbb{C}^{N_{t}\times 1}$ is the precoder, $x_{n} \in \mathbb{C}$ is the symbol, and $w_{n} \in \mathbb{C}$ is the additive noise in the system. The combined channel matrix across all the sub-carriers is $\bm{\Tilde{H}}$ with the dimensions of $\Tilde{N}_{c} \times N_{t}$.
\begin{equation}
    \bm{\Tilde{H}} = \big[\bm{\Tilde{h}_{1},\Tilde{h}_{2},\cdots,\Tilde{h}_{\Tilde{N_{c}}}} \big]^{H}.
\end{equation}
This channel matrix has a total of $2N_{t}\Tilde{N}_{C}$ feedback elements, considering the entries of the matrix are complex. In reality, transmitting such a large matrix is impractical for a massive MIMO system. Therefore, we minimize the payload by transforming the channel into the angular-delay domain which makes the channel sparser because, in this domain, multipath arrivals lie within a certain time period forcing the remaining entries close to zero \cite{csinet}. 
\begin{equation}
    \bm{\bar{H}} = \bm{F_{d}\Tilde{H}F_{a}^{H}},
\end{equation}
where $\bm{F_{d}}$ and $\bm{F_{a}}$ are 2-D discrete Fourier transform matrices of size $\Tilde{N}_{c}\times \Tilde{N}_{c}$ and $N_{t} \times N_{t}$. We then shorten the sparsified matrix $\bm{\bar{H}}$ by retaining the first $N_{c}$ non-zero rows, reducing the dimensions of the channel matrix to $N_{c}\times N_{t}$. This makes the feedback payload $2N_{c}N_{t}$ that is smaller in size and suitable for transmission. We separate the real and imaginary parts of this truncated matrix, $\bm{H}$, and treat them as two separate channels.
\par
\begin{figure}[h]
\centering
\includegraphics[width=3in]{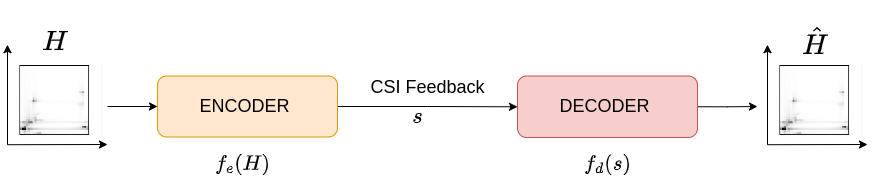}
\captionof{figure}{Auto-encoder model of CSI feedback.}
\label{net}
\end{figure}

We then compress the matrix, $\bm{H}$, into a one-dimensional vector of size $M \times 1$ as shown in Fig.{~\ref{net}}. Here, we define the compression ratio as $\gamma = \frac{M}{2N_{c}N_{t}}$. This compressed 1-D vector is then transmitted to gNB which then decodes it as $\bm{\hat{H}}$. This entire flow can be denoted by
\[s = f_{e}(\bm{H}),\] \[\bm{\hat{H}} = f_{d}(s),\]
where $f_{e}$ is the encoder, and $f_{d}$ is the decoder functions. $s$, and $\bm{\hat{H}}$ are the compressed and the estimated channel matrices respectively.

\section{Multi-Task learning}
Multi-task learning is defined as the training of a model with a certain number of tasks where all the tasks or a subset of them are related but not identical. The relatedness between the tasks might have a constructive effect on the model improving each others' performance. This relatedness is a key factor in understanding the benefits of MTL. Users distributed according to SSSM or SSMM are highly correlated as opposed to the users corresponding to MSSM or MSMM because users experiencing similar channel conditions have similar encoder outputs. This also makes them easier to learn jointly by gNB's decoder. Task relatedness gives rise to two types of MTL models that we investigate in this study: architecture-based MTL model and training-based MTL model. Architecture-based MTL models accommodate special structures in the model itself to learn the task relatedness. This can be mainly divided into two types: hard parameter sharing and soft parameter sharing \cite{overview}. On the other hand, Training based MTL models don't have any special structures to learn the task relatedness but they aim to encode it into the learning model itself via regularization \cite{types}. This can be achieved by joint training as described in the next subsection.

\subsection{Training-based MTL}
\begin{figure}[h]
\centering
\includegraphics[width=0.85\linewidth]{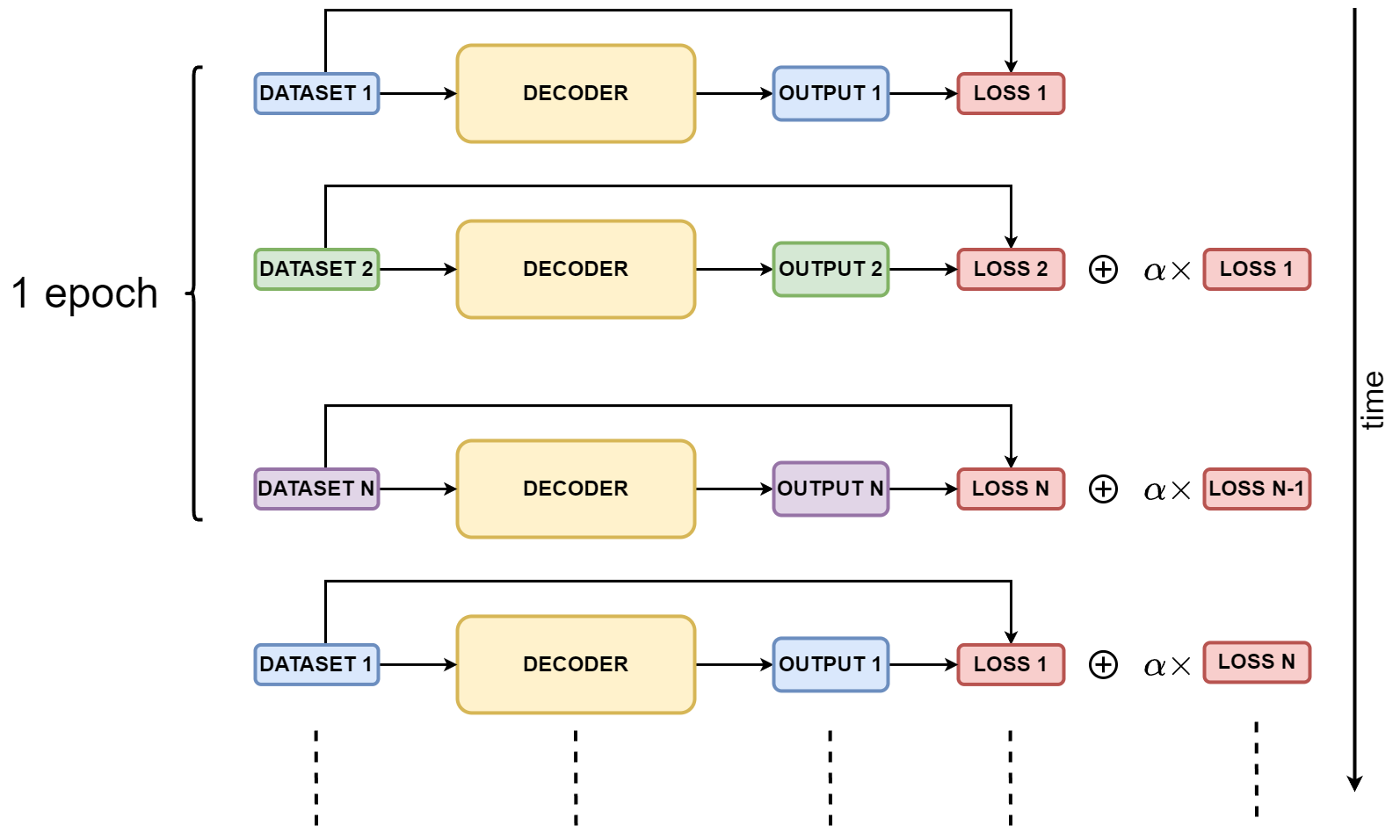}
\captionof{figure}{Joint training with the linear autoregressive loss for N tasks where each task is trained separately on the decoder by choosing a small batch size. Autoregression is achieved by adding the previous loss to the current loss.}
\label{prev}
\end{figure}

\subsubsection{Joint Training}
In every epoch, a small batch from the dataset of each task is trained on the model sequentially with all the tasks while calculating the loss for each task separately. After calculating the loss, we also add the loss from the previous training instance to the current training instance as shown in Fig.{~\ref{prev}}. So for dataset $i$, the loss function is defined as,
\begin{equation}
    loss\{i\} = loss\{i\} + \alpha \times loss\{i-1\}.
\end{equation}
Note that $i$ here means $i\pmod{N}$ as the training is cyclic but it's omitted in the equations hereafter for simplicity. This is a linear autoregressive model for the loss function where $\alpha$ is a regularization coefficient. With this, the model not only tries to learn the features of the current instance but also tries to keep its knowledge of the features from the previous instance. This facilitates the model to learn task relatedness making it effective in SSSM or SSMM user distributions as they have higher similarity between them. As the decoder in Fig.~\ref{prev} should learn the task relatedness from many encoder models, we propose to use a high-performance model like STNet\cite{stnet} as the decoder.

\subsection{Architecture-based MTL}

\subsubsection{Hard Parameter Sharing}
\begin{figure}[h]
\centering
\includegraphics[width=2in]{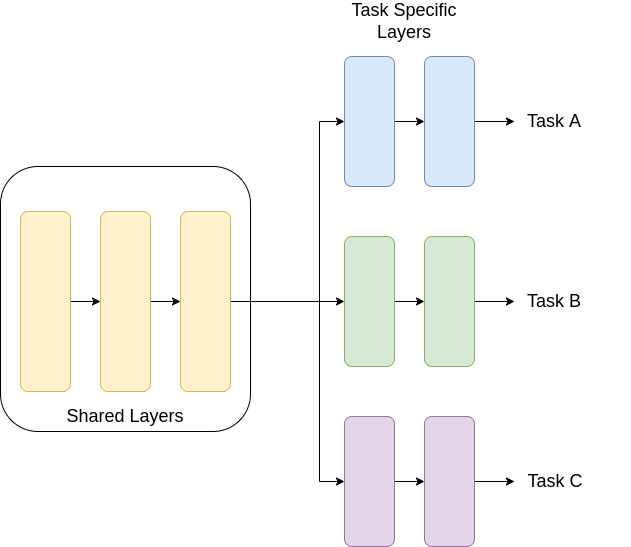}
\captionof{figure}{Decoder architecture with hard parameter sharing method to capture task relatedness.}
\label{hard}
\end{figure}

Task relatedness is captured by sharing a few layers across tasks while keeping several task-specific output layers as shown in Fig.{~\ref{hard}}.
This method also greatly reduces the risk of overfitting as it adds inductive bias by providing an alternate hypothesis for the model to learn \cite{overview} which generalizes the solutions better.
In the context of CSI feedback, soft parameter sharing is not preferred as it conditions parameters across users, which is computationally more complex so it is outside the scope of this work. In particular, for MSSM and MSMM distributions, the hard-sharing of parameters is effective as each task-specific layer can be assigned to a different channel scenario. Also, multi-scenario distributions don't share similarities between the tasks which makes it difficult to learn the tasks simultaneously using joint training.\par
\begin{figure}[h]
\centering
   \includegraphics[width=0.55\linewidth]{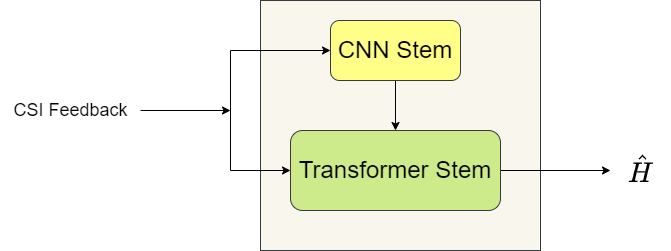}
   \caption{Structure of STNet's decoder. It consists of two stems in parallel: CNN and transformer. The output of the CNN stem is used by the transformer stem to better reconstruct the channel.}
   \label{stnet_decoder} 
\end{figure}

We propose an approach for the hard sharing of parameters exploiting the decoder architecture of STNet \cite{stnet}. Fig.~\ref{stnet_decoder} depicts the decoder of STNet consisting of two parallel stems; CNN stem and transformer stem each designed for a specific task. The transformer stem captures the long-range correlation of the antennas to better reconstruct the channel matrix whereas the CNN stem supplements it by capturing the short-range details of the channel.

\begin{figure}[h]
\centering
   \includegraphics[width=0.6\linewidth]{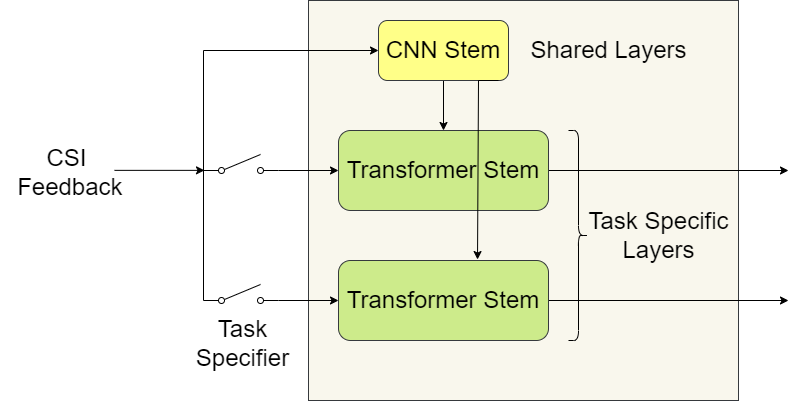}
   \caption{Proposed changes to the STNet decoder to allow the hard-sharing of parameters for two tasks. CNN stem is shared across the two tasks while the transformer stem is task specific. }
   \label{mssm} 
\end{figure}

As the transformer stem is the main contributor to the performance of the STNet, we make it task-specific while sharing the CNN stem across all tasks as shown in Fig.~\ref{mssm}. A task specifier should also be sent alongside CSI feedback to indicate which transformer stem to use. Obtaining this task specifier is outside the scope of this work, but a simple approach can be found in \cite{past}.
\section{Results}
For simulation purposes, we choose an antenna configuration of $32\times1$ antennas (i.e., gNB has 32 antennas and UE has 1 antenna). We evaluate the models on the COST2100 dataset with two scenarios: the indoor pico cellular scenario at 5.3GHz and the outdoor rural scenario at 300MHz. The training datasets consist of 100,000 matrices in each scenario whereas the validation and test datasets consist are 30,000, and 20,000 respectively. We employ mean squared error (MSE) with an Adam optimizer as the loss function.
\begin{equation}
    MSE = \frac{1}{B}\sum_{i=1}^{B} ||\bm{H}-\bm{\hat{H}}||^2,
\end{equation}
where $B$ is the batch size, $\bm{H}$, and $\bm{\hat{H}}$ are the truncated channel matrix and the decompressed channel matrix respectively from Fig.~\ref{net}. The key performance indicator is the normalized Mean Square Error (NMSE) 
\begin{equation}
    NMSE = \mathbb{E} \bigg\{\frac{||\bm{H}-\bm{\hat{H}}||^2}{||\bm{H}||^2}\bigg\}.
\end{equation}

We consider two auto-encoder models for our evaluation; CSINet\cite{csinet}, and STNet\cite{stnet}. However, the methodology can be extended to any number of models. The original NMSE values of these models when trained without any multi-task learning techniques are given in Table III. We will use these values to benchmark the MTL methods namely joint training and hard parameter sharing.
\begin{table}[h!]
\caption{Original NMSE values of CSINet and STNet}
\centering
\begin{tabular}{|c|c|c|c|c|c|} 
 \hline
Model \& Scenario & \multicolumn{1}{c|}{1/4} & \multicolumn{1}{c|}{1/16} & \multicolumn{1}{c|}{1/32} & \multicolumn{1}{c|}{1/64}\\
 \hline
 CSINet Indoor & -17.36 & -8.65& -6.24 &-5.84 \\ 
 \hline
  CSINet Outdoor & -8.75 & -4.51&  -2.81 &-1.93  \\
 \hline
 STNet Indoor & -31.81& -15.43&  -9.42 &-7.81  \\
 \hline
 STNet Outdoor & -12.91& -5.72&  -3.51 &-2.46  \\
 \hline
\end{tabular}\\
\end{table}

\subsection{Joint Training}
Joint training is advantageous in SSSM and SSMM scenarios due to the higher similarity between the tasks, so for this evaluation, we chose an indoor scenario with two encoder models: CSINet and STNet. We then trained these models on the STNet's decoder as per Fig.~\ref{prev}. The batch size was set to 50 and the regularization coefficient to 0.3 with a learning rate of 0.001. The obtained NMSE results are given in Table IV.
\begin{table}[h!]
\caption{CSINet + STNet joint training}
\centering
\begin{tabular}{|c|c c|c c|} 
 \hline
  Scenario & \multicolumn{2}{c|}{Indoor}\\
 \hline
Model& CSINet & STNet\\
 \hline
 1/4 & -19.10 & -25.32\\ 
 \hline
 1/16 & -10.92 & -14.63\\
 \hline
 1/32 & -8.68 & -9.24\\
 \hline
 1/64 & -6.22 & -6.90\\
 \hline
\end{tabular}\\
\end{table}

\begin{figure}[h]
\centering
   \includegraphics[width=\linewidth]{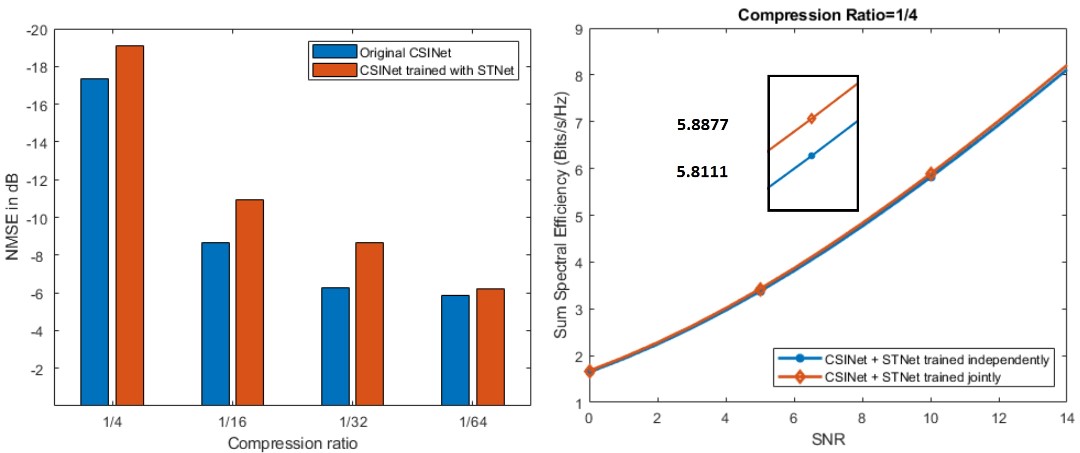}
   \caption{1) NMSE of CSINet in indoor channel scenarios when trained independently vs jointly with STNet. 2) Combined spectral efficiency plots of CSINet and STNet when trained independently vs jointly. At SNR=10dB, the values are zoomed in for clarity.}
   \label{csinet}
\end{figure}

Comparing the NMSE values of CSINet between Tables III and IV, we notice that the performance of CSINet has improved when trained along with STNet. This is presented in Fig.~\ref{csinet} for various compression ratios. However, unlike CSINet the performance of STNet decreased when trained jointly. To understand this result better, we used the linear Zero-Forcing (ZF) transmit precoding\cite{stnet} to evaluate the overall spectral efficiency improvement of the system due to joint training. Although STNet's performance has reduced, it is evident from Fig.~\ref{csinet} that the combined spectral efficiency of CSINet and STNet when trained jointly is significantly higher than the combined spectral efficiency when trained independently.

\subsection{Hard Parameter Sharing}
\begin{figure}[h]
\centering
   \includegraphics[width=\linewidth]{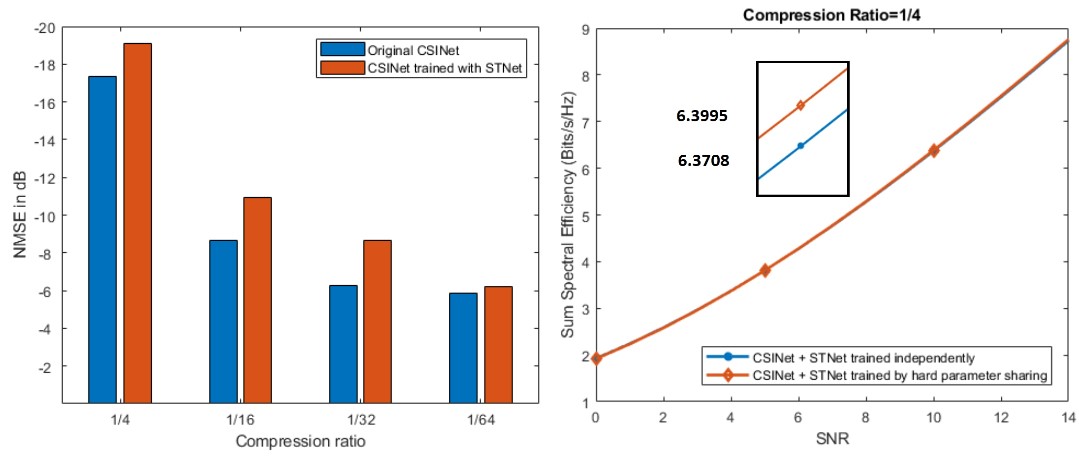}
   \caption{1) NMSE of CSINet for an indoor channel scenario when trained with STNet in an outdoor channel scenario using hard parameter sharing. 2) Combined spectral efficiency plots of CSINet and STNet when trained independently vs hard sharing of parameters. At SNR=10dB, the values are zoomed in for clarity.}
   \label{csinet_mssm} 
\end{figure}

As mentioned earlier, hard parameter sharing is efficient in MSSM and MSMM scenarios, so we chose two encoders in different scenarios: the CSINet encoder in the indoor scenario and the STNet encoder in the outdoor scenario. We then trained these models on the decoder shown in Fig.~\ref{csinet_mssm} while assigning each encoder to a different transformer stem. The batch size and learning rate are the same as earlier. The NMSE results of this configuration are given in Table V.
\begin{table}[h!]
\caption{CSINet+STNet hard parameter sharing}
\centering
\begin{tabular}{|c|c c|} 
 \hline
Model& CSINet Indoor & STNet Outdoor\\
 \hline
 1/4 & -18.08 & -10.06\\ 
 \hline
 1/16 & -11.22 & -4.92 \\
 \hline
 1/32 & -8.64 & -3.13\\
 \hline
 1/64 & -6.26 & -2.10\\
 \hline
\end{tabular}\\
\end{table}

Similar to joint training, the performance of CSINet has also improved when trained with hard parameter sharing on a modified STNet decoder. The NMSE gains of this approach over regular training are presented in Fig.~\ref{csinet_mssm}. We also used the linear Zero-Forcing (ZF) transmit precoding to gauge the spectral efficiency improvement as shown in Fig.~\ref{csinet_mssm}. 
\subsection{Complexity}
\begin{table}[h!]
\caption{Number of parameters}
\centering
\begin{tabular}{|c|c c c|} 
 \hline
  Configuration & \multicolumn{3}{c|}{CSINet+STNet}\\
 \hline
Training Type & Independent & Joint Learning & Hard Sharing \\
 \hline
 1/4 & 4.216M & 3.164M & 3.954M \\ 
 \hline
 1/16 & 1.070M & 0.804M & 0.912M \\
 \hline
 1/32 & 0.546M & 0.410M & 0.496M\\
 \hline
 1/64 & 0.283M & 0.213M & 0.252M\\
 \hline
\end{tabular}\\
\end{table}
The original purpose of joint training or hard parameter sharing is to reduce the resource utilization of the system which can be quantified by the number of parameters of the neural network. We presented the number of parameters of CSINet combined with STNet for various training methodologies in Table VI. It is evident from Table VI that the proposed methods consume significantly fewer hardware resources while achieving better spectral efficiencies. For example, CSINet in an SSMM scenario when trained jointly with STNet consumed $25\%$ fewer resources while increasing the combined spectral efficiency by $0.07bps/Hz$. Similarly, CSINet in an MSMM scenario when trained with hard parameter sharing consumed $7\%$ fewer resources while increasing its combined spectral efficiency by $0.02bps/Hz$. In addition, our approach also increased the performance of CSINet individually. For example, when trained jointly with STNet the performance of CSINet, for a compression ratio of $1/32$,  increased by $39\%$, and when trained by hard parameter sharing, it increased by $38\%$.\par Thus, by training a model in the presence of a high-performance and efficient model like STNet, we not only increased the performance of the former but also increased the throughput of the system by consuming fewer resources. 

\section{Conclusion}
In this letter, we divided the users in a multi-user scenario into four categories within the context of channel state information (CSI) feedback using deep learning and presented the inadequacies of the existing CSI feedback models in generalizing their results to all the categories. We then proposed multi-task learning techniques to address the shortcomings of these existing models to improve their performance and resource utilization in a multi-user scenario.



\bibliography{reference}
\bibliographystyle{ieeetr}
\end{document}